\documentclass[preprint,12pt]{elsarticle}

\usepackage{hyperref}

\usepackage{amssymb}
 \usepackage{amsthm}

\usepackage{mathtools, amsmath, amsbsy, amsfonts, amscd}
\usepackage{euscript}
\usepackage{bm}
\usepackage{breqn}
\usepackage{xcolor, soul}
\usepackage{empheq}
\usepackage{graphicx}
\usepackage{pstool}

\numberwithin{equation}{section}

\definecolor{III1}{rgb}{0.651, 0.495, 0.925}
\definecolor{III2}{rgb}{0, 0.615, 0}

\journal{JMPS}

\begin{document}

\begin{frontmatter}



\title{On Local Kirigami Mechanics I: Isometric Conical Solutions
\footnote{Link to the formal publication: \url{https://doi.org/10.1016/j.jmps.2021.104370}}
\footnote{\textcopyright~2021. This manuscript version is made available under the CC-BY-NC-ND 4.0 license \url{http://creativecommons.org/licenses/by-nc-nd/4.0/}}
}


\author[1]{Souhayl Sadik}
\ead{souhayl.sadik@mpe.au.dk}
\author[1,2]{Marcelo A. Dias\corref{cor1}}
\ead{marcelo.dias@ed.ac.uk}

\cortext[cor1]{Corresponding author}
\address[1]{Department of Mechanical and Production Engineering, Aarhus University, 8000~Aarhus~C, Denmark}
\address[2]{Institute for Infrastructure \& Environment, School of Engineering, The University of Edinburgh, Edinburgh~EH9~3FG Scotland, UK}

\begin{abstract}
Over the past decade, kirigami\textemdash the Japanese art of paper cutting\textemdash has been playing an increasing role in the emerging field of mechanical metamaterials and a myriad of other mechanical applications. Nonetheless, a deep understanding of the mathematics and mechanics of kirigami structures is yet to be achieved in order to unlock their full potential to pioneer more advanced applications in the field. In this work, we study the most fundamental geometric building block of kirigami: a thin sheet with a single cut. We consider a reduced two-dimensional plate model of a circular thin disk with a radial slit and investigate its deformation following the opening of the slit and the rotation of its lips. In the isometric limit\textemdash as the thickness of the disk approaches zero\textemdash the elastic energy has no stretching contribution and the thin sheet takes a conical shape known as the e-cone. We solve the post-buckling problem for the e-cone in the geometrically nonlinear setting assuming a Saint Venant-Kirchhoff constitutive plate model; we find closed-form expressions for the stress fields and show the geometry of the e-cone to be governed by the spherical elastica problem. This allows us to fully map out the space of solutions and investigate the stability of the post-buckled e-cone problem assuming mirror symmetric boundary conditions on the rotation of the lips on the open slit.
\end{abstract}




\date{24 February 2021}

\end{frontmatter}


\section{Introduction}
\label{sec.intro}



\begin{figure}
\centering
\def\svgwidth{.75\textwidth}
\begingroup%
  \makeatletter%
  \providecommand\color[2][]{%
    \errmessage{(Inkscape) Color is used for the text in Inkscape, but the package 'color.sty' is not loaded}%
    \renewcommand\color[2][]{}%
  }%
  \providecommand\transparent[1]{%
    \errmessage{(Inkscape) Transparency is used (non-zero) for the text in Inkscape, but the package 'transparent.sty' is not loaded}%
    \renewcommand\transparent[1]{}%
  }%
  \providecommand\rotatebox[2]{#2}%
  \newcommand*\fsize{\dimexpr\f@size pt\relax}%
  \newcommand*\lineheight[1]{\fontsize{\fsize}{#1\fsize}\selectfont}%
  \ifx\svgwidth\undefined%
    \setlength{\unitlength}{572.54102812bp}%
    \ifx\svgscale\undefined%
      \relax%
    \else%
      \setlength{\unitlength}{\unitlength * \real{\svgscale}}%
    \fi%
  \else%
    \setlength{\unitlength}{\svgwidth}%
  \fi%
  \global\let\svgwidth\undefined%
  \global\let\svgscale\undefined%
  \makeatother%
  \begin{picture}(1,0.54360557)%
    \lineheight{1}%
    \setlength\tabcolsep{0pt}%
    \put(0,0){\includegraphics[width=\unitlength,page=1]{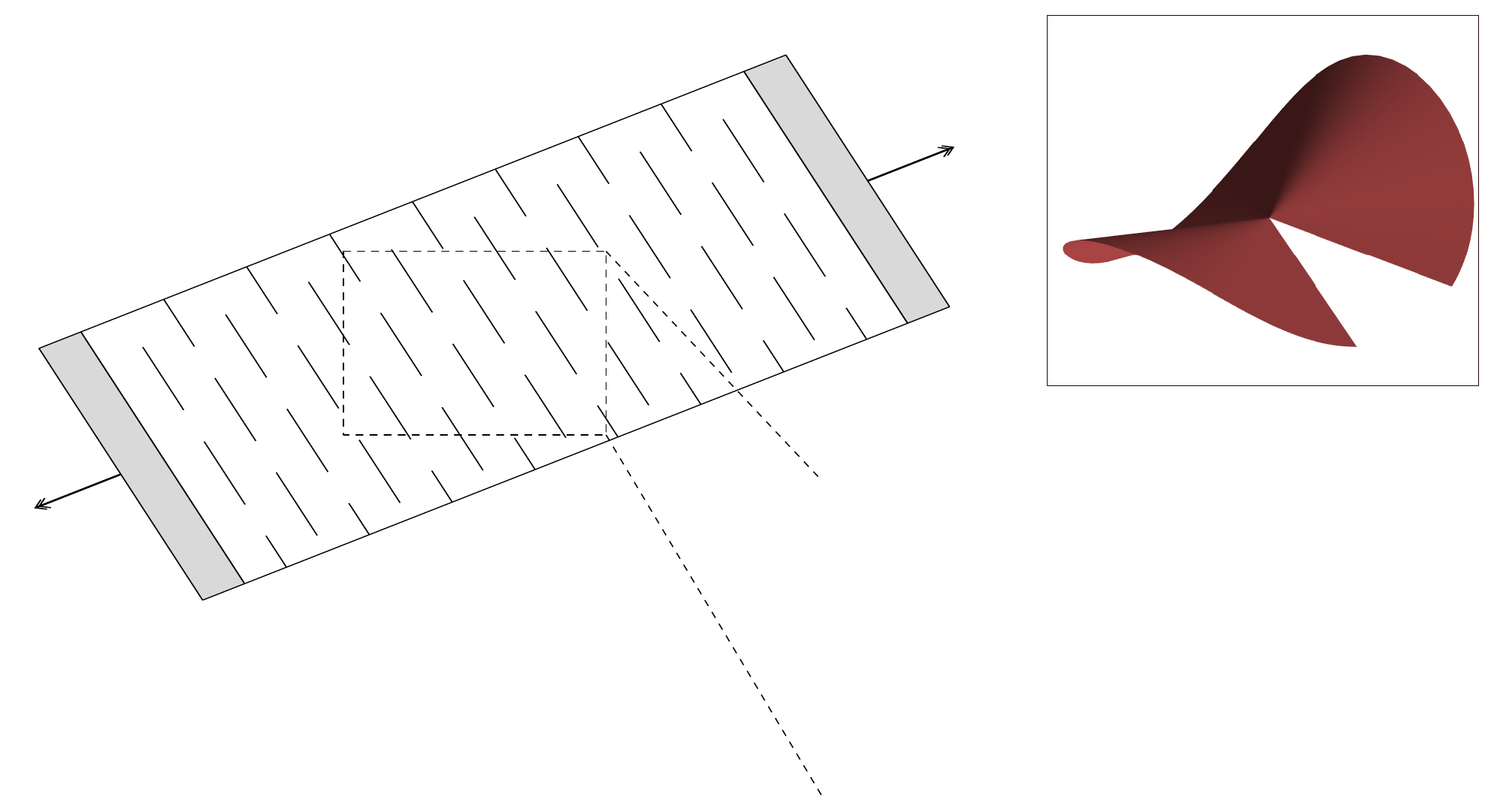}}%
    \put(0.56069137,0.46517519){\color[rgb]{0,0,0}\makebox(0,0)[lt]{\lineheight{1.25}\smash{\begin{tabular}[t]{l}\footnotesize{Stretch}\end{tabular}}}}%
    \put(-0.03508674,0.16465315){\color[rgb]{0,0,0}\makebox(0,0)[lt]{\lineheight{1.25}\smash{\begin{tabular}[t]{l}\footnotesize{Stretch}\end{tabular}}}}%
    \put(0.0204472,0.34449328){\color[rgb]{0,0,0}\makebox(0,0)[lt]{\lineheight{1.25}\smash{\begin{tabular}[t]{l}\footnotesize{(a)}\end{tabular}}}}%
    \put(0.55195008,0.2411495){\color[rgb]{0,0,0}\makebox(0,0)[lt]{\lineheight{1.25}\smash{\begin{tabular}[t]{l}\footnotesize{(b)}\end{tabular}}}}%
    \put(0.70616157,0.49272727){\color[rgb]{0,0,0}\makebox(0,0)[lt]{\lineheight{1.25}\smash{\begin{tabular}[t]{l}\footnotesize{(c)}\end{tabular}}}}%
    \put(0,0){\includegraphics[width=\unitlength,page=2]{kiri2e-c.pdf}}%
  \end{picture}%
\endgroup%

\caption{Introducing cuts in a flat sheet, i.e., creating a kirigami (a), yields a highly stretchable structure following the cuts' opening and the resulting local out-of-plane buckling (b). Locally, around the cut tip, the e-cone (c) emerges as the basic building block of kirigami.}
\label{kiri2e-c}
\end{figure}

Careful tailoring of micro-architectures in thin elastic sheets results in macroscopic structures that often reveal interesting non-linear responses to external stimuli. This is of particular interest for mechanical metamaterials, morphing structures, highly stretchable devices, and mechanical actuators. In order to illustrate this idea, let us first consider a flat sheet of paper. Depending on how it is probed, one would observe different mechanical behaviours: it is flexible when bent and stiff when subjected to an in-plane stretch. This occurs because bending offers less resistance than stretching by a factor of $(h/R)^2$, where $h$ is the thickness (smallest relevant length scale) and $R$ is a large dimension of the paper sheet. A large stretching modulus offers a high energetic cost to in-plane strains; and upon insisting on the imposition of a stretch, tearing of the sheet may occur instead. Surprisingly, by purposely placing cuts in strategic places on the sheet, as shown in Fig.~\ref{kiri2e-c}-(a), high stretchability becomes possible, thus effectively reducing the homogenized stretching modulus and consequently preventing tearing. This strategy may seem counter-intuitive, as the insertion of cuts or cracks in any medium is often associated to the introduction of potential sources of failure. However, the non-linear response to the applied stretch is such that the localized deformation at the crack tip is compensated by large out-of-plane deflections, which, due to the fact that the sheet is assumed to be thin enough, causes bending modes to be triggered much before crack propagation.
What is then observed is a beautiful textured surface, as shown in Fig.~\ref{kiri2e-c}-(b), with interesting mechanical properties. This is akin to the well-known Japanese art of paper cutting, namely kirigami, found ubiquitously in children's pop-up books. What we may call ``kirigami mechanics'' provides us with a radical paradigm shift, whereby potential sources of failure are instead functionalised. This, in turn, offers novel routes to tailor the elastic response as well as geometric pattern formation in thin elastic sheets.

Engineering applications of kirigami-based materials have recently emerged in a broad range of length-scales. From large scale morphing and deployable structures~\citep{Saito2011,Lamoureux2015} to the stretchability of graphene sheets at the atomistic level~\citep{Qi2014, Blees2015, Han2017}, kirigami has been found to be a useful tool to prescribe and control geometric features~\citep{Castle2014,Sussman2015,celli2018shape} as well as mechanical behaviours~\citep{Chen2016}. These ideas have inspired further development in stretchable electronics~\citep{Zhang2015,Song2015}, nanocomposites~\citep{Shyu2015,Xu2016}, MEMS devices~\citep{Rogers2016, Baldwin2017}, and tunable tribological properties~\citep{rafsanjani2018kirigami}. As for the multi-scale nature of the problem, it has been demonstrated, both theoretically and experimentally, that a robust link in behaviour exists across length-scales ranging over six orders of magnitude~\citep{Dias2017b}. This has made it possible to design kirigami actuators for non-linear control-response relationships capable of reliably delivering predictable motions in 3D space across multiple scales~\citep{Dias2017b,kaspersen2019lifting}. Besides these discoveries in the domain of shape-changing structures, kirigami-inspired mechanical metamaterials have been proposed for high stretchability~\citep{isobe2019continuity} and precise manipulation of materials' stiffness~\citep{dias2018}. Other types of unconventional mechanical properties have been achieved with kirigami---to name a few: zero and negative stiffness for energy dissipation~\citep{Virk2013}, auxetic behaviour~\citep{Scarpa2013,Cai2016,Tang2017}, and propagating instabilities~\citep{rafsanjani2019propagation}. 

The system shown in Fig.~\ref{kiri2e-c} demonstrates the fundamental local mechanism of interest in this article: high stretchability in kirigami is a direct consequence of the local large out-of-plane deflections induced through cut opening. In other words, we observe buckling under tension, which occurs due to a build-up of compressive stresses near the crack tip~\citep{mahmood2018cracks}. Despite recent attention given to kirigami-inspired materials, so far, analytical models capturing this rich non-linear mechanics are few and far between. Scaling laws have been derived for the amount of out-of-plane deflection near a cut, whereby an imposed tension field in the medium causes the system to equilibrate by balancing stresses at the crack tips and the available bending modes~\citep{Dias2017b}. This calculation has demonstrated that the detailed mechanics of the crack tip cannot, therefore, be neglected if we wish to understand such phenomena. Around the crack tip, stresses develop owing to an applied mode type (or a combination of the three available modes), which are related to out-of-plane bending and twisting in a non-linear way~\citep{Hui1998,Zehnder2005}. More progress has been possible when looking at the detailed geometry of the internal boundaries of cuts, which when subjected to an applied load may either tend to overlap or splay apart. It has been demonstrated that these basic motifs can be interpreted as disclinations in the medium~\citep{Castle2014}, hence locally prescribing the geometry of the entire sheet through localized sources of Gaussian curvature. Notice that, in Fig.~\ref{kiri2e-c}-(c), when the boundaries of a cut move apart from each other, an excess angle around the crack tip gives rise to negative disclination. This motif is commonly referred to as an e-cone~\citep{muller2008conical,Guven2013,efrati2015confined}, where `e' stands for \emph{excess} angle. These disclinations have strengths that are dependent on the external applied load and act as sources or charges for the system's Airy stress potential. This behaviour has been shown to be a mechanism of stress relief through a buckling response~\citep{moshe2019kirigami}. Of further interest to the scope of this article, we hope to shed light on the process that allows for precise manipulation of materials' stiffness along with the ability to tune such properties \emph{in situ}, which has been demonstrated by locally pre-programming bi-stable unit cell configurations~\citep{dias2018}.

In this article, we study the nonlinear mechanics of the e-cone\textemdash the basic building block for kirigami-based metamaterials. We propose an analytical model to resolve the geometry and mechanics around the tip of the cut in the regime of large out-of-plane deflections\textemdash when geometric non-linearities play a critical role\textemdash in the isometric limit. We derive general and explicit expressions for the e-cone shape as well as the stress fields near the cut region. Further, we focus on the effects that careful control of the edges of the cut, namely lips, has on questions of stability. By opening the slit and subjecting its lips to a mirror symmetric imposed angular distribution, we are able to track the whole space of solutions and give a complete picture of the stability map of e-cones. We are, hence, providing a solution to the local post-buckling problem in kirigami. In a separate work \cite{SaDias2021CreaKiri}, using a linearised discretely creased model for the e-cone motif, we are able to relax the inextensibility constraint\textemdash hence including the stretching energy contribution, which allows us to study the onset of buckling and post-buckling behaviour.

The manuscript is organised as follows.
In \S\ref{kinematics}, we derive the kinematics of e-cones as we lay out a geometric description of the deformation of the idealised thin sheet as a plate. The model hinges on the assumption of developability and further ensuring inextensibility.
In \S\ref{mechanics}, assuming a Saint Venant-Kirchhoff constitutive model for plates, we derive the balance laws and find a closed-form solution for the stress field in the e-cone. We also find the governing equation for the shape of the e-cone as the solution of the spherical elastica problem in terms of the normal curvature of the e-cone.
In \S\ref{results}, we specialise to the case of mirror symmetric boundary conditions on the lips rotation following the opening of the slit. We are able to map out the full space of post-buckling solutions as well as investigate their orbits' stability.
In \S\ref{conclusion}, we provide concluding remarks.

\section{Kinematics of e-cones}
\label{kinematics}

\begin{figure}
\centering
\def\svgwidth{.75\textwidth}
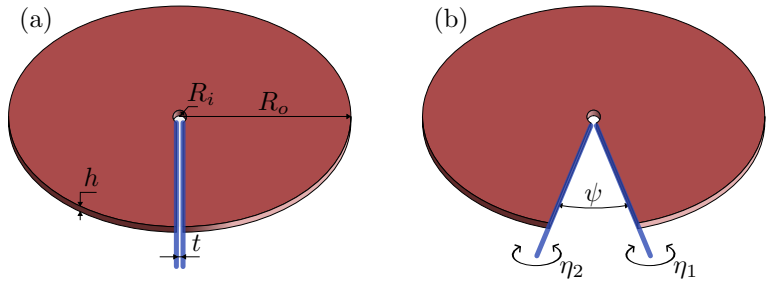
\caption{Geometry of the disk and its loading parameters. (a) The disk in its relaxed configuration. It has an inner radius $R_i\,$, an outer radius $R_o\,$, thickness $h\,$, and a slit to the centre of width $t\,$.
(b) The disk is loaded by an in-plane rotation of the lips around its centre in order to control $\psi\,$, the opening angle of the slit; and by rotating the lips around their respective axes to control their rotation angles $\eta_1$ and $\eta_2\,$, respectively.}
\label{geometry}
\end{figure}

\begin{figure}[t]
\centering
\def\svgwidth{.75\textwidth}
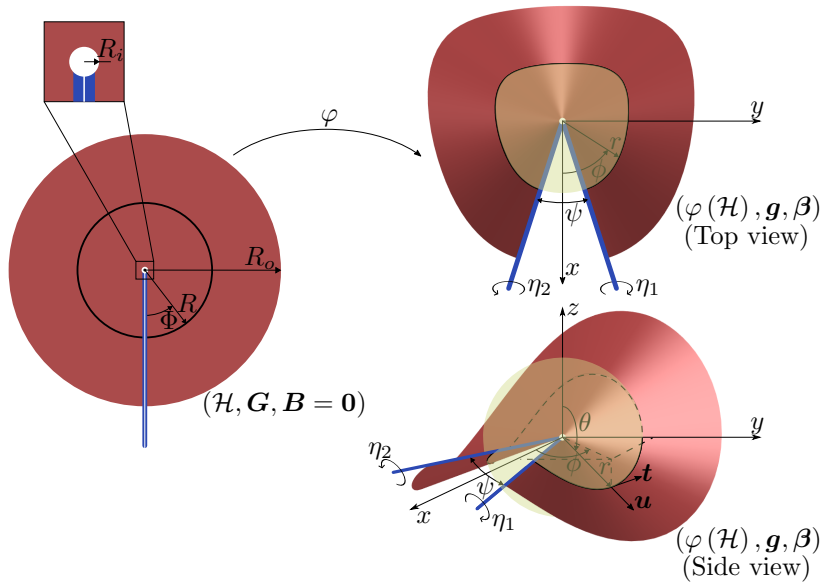
\caption{The plate model $\left(\mathcal H,\boldsymbol G,\boldsymbol B=\boldsymbol 0\right)$ of the disk deforms in the ambient space $\mathcal S$ via a deformation mapping $\varphi:\mathcal H \to \mathcal S$ to form a buckled configuration $\left(\varphi\left(\mathcal H\right),\boldsymbol g,\boldsymbol \beta\right)\,$.
The conical ansatz given in Eq.~\eqref{ansatz} is such that a circle of radius $R$ centred at the material origin on the undeformed plate deforms into a curve that lives on the translucent sphere of the same radius $r=R$ centred at the spatial origin in the deformed configuration. $\boldsymbol t$ and $\boldsymbol u$ are the unit tangent and the unit in-plane outward normal of such a curve, respectively.
}
\label{not_potato}
\end{figure}

We consider a thin, initially planar, and circular disk\textemdash or rather annulus\textemdash of thickness $h\,$, outer radius $R_o\,$, and inner radius $R_i\,$, with a radial thin slit of width $t$ to the centre; see Fig.~\ref{geometry}-(b). The hole in the centre is made so as to prevent a logarithmic singular bending energy at the apex\textemdash as may be seen from \eqref{tot_E}, if $R_i$ were to be taken to the zero limit. Indeed, in engineering applications, the kirigami slits are typically made by laser cutting and hence are such that $R_i\neq 0$ and $t\neq 0\,$, which helps with reducing stress concentration at the apex.
We concern ourselves with the study of the deformation of this structure following the opening of the slit with an angle $\psi$ and the rotation of its lips with angles $\eta_1$ and $\eta_2\,$, respectively.
The slit opens such that its lips rotate around the centre of the disk to form an angle $\psi$ while remaining on the original plane of the disk's mid-surface.
See Fig.~\ref{geometry} for a schematic of the geometry of the disk and how it is loaded.\footnote{Note that for the sake of simplicity of presentation at this point of the manuscript, Fig.~\ref{geometry} does not depict the out-of-plane deformation of the disk. Such a depiction may however be found in Fig.~\ref{not_potato} and later on in the results' section, \S \ref{results}.}

In its undeformed state, we reduce the disk to its mid-surface and identify it with a two-dimensional circular plate $\mathcal H$ sitting in $\mathbb R^3$. Further, the slit of the disk is modelled by a radial width-less cut to the centre.
We let $\left\{R,\Phi\right\}$, $0\leq\Phi\leq2\pi\,$, be a polar coordinates system on $\mathcal H$ such that its origin is located at the centre of $\mathcal H\,$; and such that $\Phi=0$ and $\Phi=2\pi$ arbitrarily and respectively correspond to the two lips of the slit.
As a planar hyper-surface in $\mathbb R^3\,$, $\mathcal H$ is equipped with a first fundamental form $\boldsymbol G$ which in local Lagrangian coordinates $\left\{R,\Phi\right\}$ reads as $\boldsymbol G = \operatorname{diag}\left(1,R^2\right)\,$; and with a vanishing second fundamental form $\boldsymbol B = \boldsymbol 0\,$. We identify the three-dimensional ambient space $\mathcal S$ with $\mathbb R^3$ and denote by $\bar{\boldsymbol g}$ its flat Euclidean metric and by $\nabla^{\bar{\boldsymbol g}}$ its Levi-Civita connection.
We let $\left\{r,\theta,\phi\right\}$, $0\leq\theta\leq\pi\,$, $0\leq\phi<2\pi$ be a spherical coordinate system in $\mathbb R^3\,$.
The deformation of the disk into the ambient space is modelled by a deformation mapping $\varphi:\mathcal H \to \mathcal S$. See Fig.~\ref{not_potato} for a schematic representing the plate model and its deformation.
We adopt the standard convention to denote objects and indices in the material (reference) manifold $\mathcal H$ by uppercase characters (e.g., a material point $X \in \mathcal H$) and in the spatial (deformed) manifold $\varphi(\mathcal H) \subset \mathbb R^3$ by lowercase characters (e.g., a spatial point $x \in \varphi(\mathcal H)$). Unless otherwise stated, we adopt Einstein's repeated index summation convention, i.e., $T^A{}_A = \sum_{K=1}^2 T^K{}_K$ and $t^a{}_a = \sum_{k=1}^3 t^k{}_k\,$.

For small values of $\psi\,$, the plate undergoes circumferentially-uniform compression and deforms into a planar configuration\textemdash as depicted in Fig.~\ref{geometry}-(b). However, beyond a certain critical value $\psi=\psi_c$, this planar configuration becomes unstable and buckles out-of-plane\textemdash as depicted in Fig.~\ref{not_potato}. Similarly to a rectangular plate under axial compression \citep{audoly2010elasticity}, the critical out-of-plane buckling threshold for a disk of radius $R_0$ scales as $\psi_c \propto (h/R_0)^2\,$.
In this work, we are interested in studying thin sheets in their isometric limit, i.e., as the thickness of the thin sheet approaches zero ($h/R_0\ll1\,$) and deforms isometrically.
Note that the isometric deformation of the plate equates to assuming inextensibility and yields that the plate should buckle out-of-plane at $\psi_c=0\,$\textemdash because isometry dictates that the deformation does not induce any in-plane stretch.
Indeed, it would later become apparent from the energy argument presented in the beginning of \S\ref{mechanics} why we may assume inextensibility and $\psi_c=0$ as we approach the isometric limit.
Besides, we are not interested in this work to explore the onset of the e-cone out-of-plane instability in itself; instead, we are interested in studying the buckled configuration that follows it and its post-buckling stability.
Therefore, in what follows, we assume that the plate deforms isometrically; which, as noted earlier, equates to assuming an inextensible plate indeed and ensures that it remains developable after deformation.

As the slit of the flat sheet opens, the only possibility for the sheet to remain developable and maintain its structural integrity (i.e., no crack propagation) is for its radial generators to remain straight and unstretched; which only ensures radial inextensibility, we will later make sure the deformation is indeed an isometry. In terms of the deformation embedding, fixing the origin of $\left(r,\theta,\phi\right)$ at the centre of the plate, it amounts to assuming the following conical ansatz for the deformation mapping
\begin{equation}
\label{ansatz}
\varphi(R,\Phi)=\left(R,\theta(\Phi),\phi(\Phi)\right)\,.
\end{equation}
We further assume the following boundary conditions to enforce the opening of the slit by an angle $\psi$ {while keeping its lips in the disk's initial plane} and control {their respective} angles of rotation by $\eta_1$ and $\eta_2\,$
\begin{equation}\label{ansatz_bdcd}
    \begin{gathered}
        \phi(0) = \frac{\psi}{2} \,, \quad \phi(2\pi) = 2\pi - \frac{\psi}{2} \,, \quad \theta(0) = \theta(2\pi) = \frac{\pi}{2} \,, \\
	{\sin\left[\theta(0)\right]\theta'(0) = \sin(\eta_1)} \,, \quad {\sin\left[\theta(2\pi)\right]\theta'(2\pi) = \sin(\eta_2)} \,.
    \end{gathered}
\end{equation}
We also assume traction-free, shear-free, and moment-free boundary conditions on the inner and outer boundaries of the plate, i.e., at $R=R_o$ and $R=R_i\,$, respectively.
Introducing an excess angle $\psi$ to constrain the conical shape of the buckled plate given by the conical ansatz in Eq.~\eqref{ansatz} leads to the so-called e-cone shape. Note that the conical ansatz is such that a circle of radius $R$ centred at the material origin on the undeformed plate deforms into a curve that lives on a sphere of the same radius $r=R$ centred at the spatial origin on the deformed configuration\textemdash see Fig.~\ref{not_potato}. A parametrization of such a curve is exactly given by Eq.~\eqref{ansatz} for fixed $R\,$, i.e., $\Phi\mapsto\left(R,\theta(\Phi),\phi(\Phi)\right)\,$.

In the deformed configuration $\varphi(\mathcal H)\subset\mathbb R^3\,$, we denote the normal of the surface by $\boldsymbol n$ and its first and second fundamental forms by $\boldsymbol g= \bar{\boldsymbol g}_{|\varphi(\mathcal H)}$ and $\boldsymbol{\beta}=-\left(\nabla^{\bar{\boldsymbol g}}\boldsymbol n\right)^\flat\,$,\footnote{The flat symbol in $(.)^\flat$ denotes the operator for lowering tensor indices.} respectively.
To quantify the strain, we define an intrinsic strain measure $\boldsymbol C = \varphi^*\boldsymbol g\,$\footnote{$\varphi^*$ denotes the pullback by the diffeomorphism $\varphi\,$. As examples, the pullback of a $({}^0_2)$-rank tensor $\boldsymbol w$ reads in local coordinates $\left(\varphi^*\boldsymbol w\right)_{AB} = \left(\partial \varphi^a/\partial X^A\right)\left(\partial \varphi^b/\partial X^B\right)w_{ab}\,$; and the pullback of a $({}^2_0)$-rank tensor $\boldsymbol \omega$ reads in local coordinates $\left(\varphi^*\boldsymbol \omega\right)^{AB} = \left(\partial \varphi^{-A}/\partial x^a\right)\left(\partial \varphi^{-B}/\partial x^b\right)\omega^{ab}\,$.}\textemdash the right Cauchy-Green deformation tensor\footnote{Note that the right Cauchy-Green deformation tensor is typically defined as $\boldsymbol F^{\mathsf T} \boldsymbol F$ where $^{\mathsf T}$ denotes the transpose operator. In our notation, we have $\boldsymbol C = \left(\boldsymbol F^{\mathsf T} \boldsymbol F\right)^\flat\,$.}\textemdash and an extrinsic strain measure $\boldsymbol \Theta = \varphi^*\boldsymbol \beta\,$.
We denote by $J$ the Jacobian of the deformation, and it can be shown that $J = \sqrt{{\det \boldsymbol C}/{\det \boldsymbol G}}\,$.
We define the convected manifold $\left(\mathcal H,\boldsymbol C,\boldsymbol \Theta\right)$ as the Riemannian manifold resulting from pulling-back the geometry of the deformed surface onto $\mathcal H\,$, i.e., taking $\boldsymbol C$ and $\boldsymbol \Theta$ to be the first and second fundamental forms of $\mathcal H\,$, respectively.
In the local chart $\{R,\Phi\}$, the components of $\boldsymbol C$ and $\boldsymbol \Theta$ read as follow
\begin{equation}
\label{fund_forms_coord}
C_{AB}=\varphi_{,A}\cdot\varphi_{,B}\,,\quad
\Theta_{AB}=\varphi_{,AB}\cdot\frac{\varphi_{,R}\times\varphi_{,\Phi}}{\|\varphi_{,R}\times\varphi_{,\Phi}\|}\,.
\end{equation}
where $\cdot\,\,$, $\times\,\,$, and $\|.\|$ denote the dot product, the cross product, and the standard norm in $\mathbb R^3\,$, respectively. 

For the ansatz given in Eq.~\eqref{ansatz}, and following Eq.~\eqref{fund_forms_coord}, the first fundamental form reads in $\left\{R,\Phi\right\}$ as
\begin{equation}\nonumber
\boldsymbol C = \left(
\begin{array}{cc}
 1 & 0 \\
 0 & R^2 \left[{\theta'}^2(\Phi )+\sin ^2(\theta (\Phi )) {\phi'}^2(\Phi )\right] \\
\end{array}
\right)\,.
\end{equation}
To complete enforcing the inextensibility condition, which amounts to assuming the deformation map to be an isometry, i.e., $\boldsymbol C = \boldsymbol G\,$,\footnote{Note that the inextensibility condition is stronger than the incompressibility condition. Incompressibility requires that the volume is preserved, i.e., $J=1$; while inextensibility requires that the deformation does not induce any in-plane stretch of the material, i.e., $\boldsymbol C = \boldsymbol G\,$, which indeed implies that the volume is preserved.} we need to have $\theta '(\Phi )^2+\sin ^2(\theta (\Phi )) \phi '(\Phi )^2=1\,$. {
We assume that the e-cone does not fold over itself, i.e., $\phi'(\Phi)\geq 0\,$, and it hence} follows that
\begin{equation}\label{phi}
\phi(\Phi) = \frac{\psi }{2} + \int_0^{\Phi } \frac{\sqrt{1-{\theta'}^2(\zeta )}}{\sin [\theta (\zeta )]} \, d\zeta\,,
\end{equation}
where only one boundary condition\textemdash $\,\phi(0) = \psi/2\,$\textemdash out of the six appearing in Eq.~\eqref{ansatz_bdcd} has been used. Finally, using Eq.~\eqref{fund_forms_coord} and Eq.~\eqref{phi}, the second fundamental form for the assumed ansatz, i.e., Eq.~\eqref{ansatz}, reads in $\left\{R,\Phi\right\}$ as
\begin{equation}\nonumber
\boldsymbol \Theta = \left(
\begin{array}{cc}
 0 & 0 \\
 0 & R \kappa(\Phi)
\end{array}
\right)\,,
\end{equation}
where we introduced {the normal curvature}
\begin{equation}\label{kappa_def}
\kappa(\Phi):=\frac{\cot [\theta (\Phi )] \left(1-{\theta'}^2(\Phi )\right)-\theta ''(\Phi )}{\sqrt{1-{\theta'}^2(\Phi )}}\,.
\end{equation}

As noted earlier, the conical ansatz given in Eq.~\eqref{ansatz} is such that every circle of radius $R$ centred at the material origin deforms into a curve that lives on the sphere of radius $R$ centred at the spatial origin. However, the geometry of such curves is further constrained by the linear $R$-dependence of the ansatz forcing the curves to form a conical surface. As such, for a fixed radius $R$, the intrinsic curvature $\kappa_c$ of such a curve reads $\kappa_c^2(R,\Phi)=R^2(1+\kappa^2(\Phi))\,$.

Note that besides the inherent spherical polar coordinates induced indeterminacy at the origin, the assumed ansatz, i.e., Eq.~\eqref{ansatz}, violates the inextensibility condition at the apex of the e-cone, i.e., at $R=0\,$. Indeed, as discussed earlier, inextensibility dictates that the initially flat sheet should remain developable after the deformation; however, the e-cone is flat almost everywhere except on its apex at $R=0$ where there would be a concentrated curvature charge due to the excess angle~$\psi\,$.
Cutting out a hole around the apex wouldn't solve this issue \citep{yu2020cutting}; indeed, it turns out that the violation of the inextensibility condition is unavoidable. As we will later discuss following Eq.~\eqref{traction}, it would be necessary to relax the inextensibility condition on both an inner and an outer boundary layer to ensure the system's global equilibrium. 
However, cutting out a small disk of material of radius $R_i$ around the apex is still necessary to get around a bending energy singularity (and a resulting stress singularity) at the apex as $R$ approaches $0$ that would later be revealed by the expression of the elastic energy in Eq.~\eqref{tot_E} (and the stress computations in Eq.~\eqref{traction}, respectively).

\section{Constitutive behaviour and balance law}
\label{mechanics}

For a plate, the elastic energy density $\mathcal W$ may be approximated by the sum of a stretching contribution $\mathcal W_s$ and a bending contribution $\mathcal W_b\,$, i.e., $\mathcal W= \mathcal W_s + \mathcal W_b\,$, such that $\mathcal W_s \propto h$ and $\mathcal W_b \propto h^3\,$, $h$ being the thickness of the plate \citep{koiter1960consistent}. In its isometric limit, which is our scope of interest in this work, a sheet would hence rather bend than stretch as the energetic cost is 2 orders of magnitude more favourable. In this case, in-plane strains would have a much higher energetic cost and the plate would be forced to buckle as soon as the slit opens. Thus, it appears reasonable to assume an inextensible plate. Furthermore, inextensibility constrains the deformation embedding to be an isometry, which as such, preserves the Gaussian curvature. Therefore, starting with a flat sheet, the deformed configuration needs to be flat as well, i.e., a developable surface.

In what follows, we assume that the disk is made of a material with a Saint Venant-Kirchhoff constitutive model.
In the convected manifold $\left(\mathcal H,\boldsymbol C,\boldsymbol \Theta\right)\,$, the strain energy density $\mathcal W$ of the material reads\footnote{Note that in the convected manifold, raising ($(.)^\sharp$) and lowering ($(.)^\flat$) tensor indices is performed using the convected metric $\boldsymbol C\,$, e.g., $T^A{}_B=C^{AK} T_{KB}\,$; indeed, one has $C^{AK}C_{KB}=\delta^A{}_B\,$. Also, the trace of a tensor is computed by using the convected metric $\boldsymbol C\,$, e.g., $\textrm{tr}\left(\boldsymbol T\right)=C^{AB}T_{AB}$ and $\textrm{tr}\left(\boldsymbol C\right)=3\,$.}
\begin{equation}
\begin{split}
\label{S_V-K}
\mathcal W
=& \frac{h}{4}\left\{\mu\textrm{tr}\left[\left(\boldsymbol C-\boldsymbol G\right)^2\right]+\frac{\mu\lambda}{2\mu+\lambda}\left[\textrm{tr}\left(\boldsymbol C-\boldsymbol G\right)\right]^2\right\} - \boldsymbol p\!:\!\left(\boldsymbol C-\boldsymbol G\right)\\
&+\frac{h^3}{12}\left\{\mu\textrm{tr}\left[\left(\boldsymbol \Theta-\boldsymbol B\right)^2\right]+\frac{\mu\lambda}{2\mu+\lambda}\left[\textrm{tr}\left(\boldsymbol \Theta-\boldsymbol B\right)\right]^2\right\}
\,,
\end{split}
\end{equation}
where $h$ is the thickness of the disk, $\mu$ and $\lambda$ the Lam\'e coefficients, and $\boldsymbol p$ a $\left({}^2_0\right)$-rank tensor Lagrange multiplier enforcing the inextensibility condition $\boldsymbol C=\boldsymbol G\,$. Without any loss of generality, we assume the following form for the Lagrange multiplier $\boldsymbol p$
\begin{equation}\nonumber
\boldsymbol p(R,\Phi) = \left(
\begin{array}{cc}
 \scriptstyle{p_1(R,\Phi )}	& \frac{\tau (R,\Phi )}{R} \\
\vspace{-10pt}\\
 \frac{\tau (R,\Phi )}{R}		& \frac{p_2(R,\Phi )}{R^2} \\
\end{array}
\right)\,.
\end{equation}
The convected stress and couple-stress tensors\footnote{Note that the convected stress and couple-stress tensors are respectively defined as the pullback by the deformation mapping of the Cauchy stress tensor $\boldsymbol \sigma$ and the spatial couple-stress tensor $\boldsymbol \mu\,$, i.e., $\boldsymbol \Sigma := \varphi^*\boldsymbol \sigma$ and $\boldsymbol \Lambda :=  \varphi^*\boldsymbol \mu\,$.} are respectively given by $\boldsymbol \Sigma = 2/J \left(\partial \mathcal W/\partial \boldsymbol C\right)$ and $\boldsymbol \Lambda = 1/J \left(\partial \mathcal W/\partial \boldsymbol  \Theta\right)\,$. It hence follows that
\begin{equation}\nonumber
\boldsymbol \Sigma = - \left(
\begin{array}{cc}
 \scriptstyle{p_1(R,\Phi )}	& \frac{\tau (R,\Phi )}{R} \\
\vspace{-10pt}\\
 \frac{\tau (R,\Phi )}{R}			& \frac{p_2(R,\Phi )}{R^2} \\
\end{array}
\right)
\,,\quad
\boldsymbol \Lambda = \left(
\begin{array}{cc}
 \frac{h^3 \lambda  \mu  \kappa (\Phi )}{6 R (\lambda +2 \mu ) }	& 0 \\
\vspace{-10pt}\\
 0																& \frac{h^3 \mu  (\lambda +\mu ) \kappa (\Phi )}{3 R^3 (\lambda +2 \mu )} \\
\end{array}
\right)\,.
\end{equation}
Not surprisingly, note that the inextensibility condition $\boldsymbol C = \boldsymbol G$ implies that the contribution of the stretching energy vanishes, as can be observed in Eq.~\eqref{S_V-K}. The buckling of the plate is then solely governed by the bending energy; however, the plate remains indeed constrained by the inextensibility condition and the associated Lagrange multiplier $\boldsymbol p$ emerges to enforce it in the form of a in-plane stress field.

We now turn our attention to the governing equations. In the convected manifold $\left(\mathcal H,\boldsymbol C,\boldsymbol \Theta\right)\,$, the balance laws for shells read \citep{Niordson1985}
\begin{subequations}
\label{Bal_f}
\begin{align}
\operatorname{Div}_{\boldsymbol C} \left(\boldsymbol \Sigma + \boldsymbol C^\sharp \cdot \boldsymbol \Theta \cdot \boldsymbol \Lambda\right)
+ \boldsymbol C^\sharp \cdot \boldsymbol \Theta \cdot \operatorname{Div}_{\boldsymbol C} \left(\boldsymbol \Lambda\right) =& \boldsymbol 0 \,,
\\%
\left(\boldsymbol \Sigma + \boldsymbol C^\sharp \cdot \boldsymbol \Theta \cdot \boldsymbol \Lambda\right) \!:\! \boldsymbol \Theta - \Delta_{\boldsymbol C} \left( \boldsymbol \Lambda \right) =&0\,,
\end{align}
\end{subequations}
where $\operatorname{Div}_{\boldsymbol C}$ and $\Delta_{\boldsymbol C}$ respectively denote the divergence and the Laplace operators in the convected manifold. In components, $\operatorname{Div}_{\boldsymbol C}\boldsymbol\Sigma = \Sigma^{AB}{}_{||B}$ and $\Delta_{\boldsymbol C}\boldsymbol\Lambda = \Lambda^{AB}{}_{||AB}\,$, where a subscripted double stroke ``$||$'' denotes covariant differentiation with respect to $\boldsymbol C$ taken as a metric on $\mathcal H\,$. We explicitly compute the governing equations above, i.e., Eq.~\eqref{Bal_f}, and obtain the following system of partial differential equations
\begin{subequations}
\label{Bal}
\begin{align}
\label{Bal1}
{p_1(R,\Phi )} + Rp_{1,R}(R,\Phi ) - {p_2(R,\Phi )} + {\tau_{,\Phi}(R,\Phi )} + \frac{h^3 \mu  (\lambda +\mu )\kappa^2(\Phi ) }{3 R^2 (\lambda +2 \mu )} =& 0
\,,\\%
\label{Bal2}
p_{2,\Phi}(R,\Phi )+R \tau_{,R}(R,\Phi )+2  \tau (R,\Phi)-\frac{h^3 \mu  (\lambda +\mu ) \kappa (\Phi ) \kappa '(\Phi )}{R^2 (\lambda +2 \mu )} =& 0
\,,\\%
\label{Bal3}
\kappa ''(\Phi ) - \kappa^3 (\Phi ) + \left(\frac{ 3 (\lambda +2 \mu )R^2 p_2(R,\Phi )}{h^3 \mu  (\lambda +\mu)}+1\right)\kappa (\Phi ) =& 0
\,.
\end{align}
\end{subequations}

From Eq.~\eqref{Bal2}, it follows that 
$$p_2(R,\Phi) = \frac{h^3 \mu  (\lambda +\mu )\kappa^2 (\Phi) }{2 R^2 (\lambda +2 \mu )} -\int_0^{\Phi } \left(R \tau_{,R}(R,\zeta )+2 \tau (R,\zeta )\right) \, d\zeta+f(R)\,,$$
for some function $f$ that depends only on the radius $R\,$.
Now, if we look at Eq.~\eqref{Bal3}, it appears that $R^2p_2(R,\Phi)$ ought to be a function of $\Phi$ only, i.e., $\partial/\partial R\left[R^2p_2(R,\Phi)\right]=0\,$; which, by using the expression above for $p_2\,$, yields for all $0\leq\Phi\leq2\pi$
\begin{equation}\label{int_tau}
\frac{\partial}{\partial R}\left[-R^2\int_0^{\Phi } \left(R \tau_{,R}(R,\zeta )+2 \tau (R,\zeta )\right) \, d\zeta+R^2f(R)\right]=0\,.
\end{equation}
In particular, for $\Phi=0\,$, one has $\partial\left[R^2f(R)\right]/\partial R=0\,$, and we may accordingly let
$$f(R)=\frac{\alpha  h^3 \mu  (\lambda +\mu )}{3 R^2 (\lambda +2 \mu )}\,,$$
for some constant $\alpha\,$.
Using the expression above for $f\,$, and by localizing the integral term in \eqref{int_tau}, it follows that
$$\frac{\partial}{\partial R}\left[R^2\left(R \tau_{,R}(R,\Phi )+2 \tau (R,\Phi)\right)\right]=0\,,$$
which in turn yields that
$$\tau(R,\Phi) = \frac{a(\Phi)}{R^2} + \frac{\ln(R)b(\Phi)}{R^2}\,,$$
for some functions $a$ and $b$ of $\Phi\,$. Now, having closed-form expressions for both $p_2$ and $\tau\,$, we use them in Eq.~\eqref{Bal1} to solve for $p_1\,$; and we summarize the expressions for the components of $\boldsymbol p$ as follow
\begin{subequations}
\label{p}
\begin{align}
\label{p1}
p_1(R,\Phi)=&-\frac{h^3 \mu  (\lambda +\mu ) \left(\kappa (\Phi )^2 + 2 \alpha \right)}{6 R^2 (\lambda +2 \mu )}\nonumber\\
&+ \frac{1}{R^2}\left(a'(\Phi )+ (\ln (R)+1) b'(\Phi )+\int_0^{\Phi } b(\zeta ) \, d\zeta\right)
+ \frac{c(\Phi)}{R}\,,
\\
\label{p2}
p_2(R,\Phi) =& \frac{h^3 \mu  (\lambda +\mu )\left(3 \kappa^2 (\Phi) + 2\alpha\right) }{6 R^2 (\lambda +2 \mu )} -\frac{1}{R^2}\int_0^{\Phi } b(\zeta) \, d\zeta\,,
\\\label{tau}
\tau(R,\Phi) =& \frac{a(\Phi)}{R^2} + \frac{\ln(R)b(\Phi)}{R^2}\,,
\end{align}
\end{subequations}
for some constant $\alpha$ and some functions $a\,$, $b\,$, and $c$ of the angle $\Phi\,$. Using the above expressions, i.e., Eq.~\eqref{p}, the system of balance equations appearing in Eq.~\eqref{Bal} reduces to the following governing equation for $\kappa$
\begin{equation}
\label{b_elastica}
\kappa ''(\Phi ) + \frac{\kappa^3 (\Phi )}{2} + \left(1+\alpha -\frac{3 (\lambda +2 \mu ) }{h^3 \mu  (\lambda +\mu )}\int_0^{\Phi } b(\zeta ) \, d\zeta\right)\kappa (\Phi ) = 0\,,
\end{equation}
which we shall be revisiting later in this work.

Let us now explore the effective physical stress fields in the plate. Along a direction $\boldsymbol d$ in the deformed configuration of the shell, the traction vector $\boldsymbol f_{\boldsymbol d}\,$, the moment vector $\boldsymbol m_{\boldsymbol d}\,$, and the out-of-plane shear $s_{\boldsymbol d}$ are respectively given by\footnote{$\varphi_*$ denotes the pushforward by the diffeomorphism $\varphi\,$. As examples, the pushforward of a $({}^2_0)$-rank tensor $\boldsymbol \Omega$ reads in local coordinates $\left(\varphi_*\boldsymbol \Omega\right)^{ab} = \left(\partial \varphi^a/\partial X^A\right)\left(\partial \varphi^b/\partial X^B\right) \Omega^{AB}\,$; and the pushforward of a $({}^0_2)$-rank tensor $\boldsymbol W$ reads in local coordinates $\left(\varphi_*\boldsymbol W\right)_{ab} = \left(\partial \varphi^{-A}/\partial x^a\right)\left(\partial \varphi^{-B}/\partial x^b\right) W_{AB}\,$.} \citep{Niordson1985}
\begin{equation}\nonumber
\begin{split}
\boldsymbol f_{\boldsymbol d} =& \varphi_*\left[\left(\boldsymbol \Sigma + 2 \boldsymbol C^\sharp \cdot \boldsymbol \Theta \cdot \boldsymbol \Lambda\right) \cdot \boldsymbol D^{\flat} - 
\left(\boldsymbol D^{\flat} \cdot \boldsymbol \Lambda \cdot \boldsymbol D^{\flat} \right) \boldsymbol C^\sharp \cdot \boldsymbol \Theta \cdot \boldsymbol D\right] \,,\\
\boldsymbol m_{\boldsymbol d} =&  \varphi_*\left[\boldsymbol \Lambda \cdot \boldsymbol D\right]\,,\quad\quad
s_{\boldsymbol d }= -\left(\operatorname{Div}_{\boldsymbol C} \boldsymbol \Lambda\right) \cdot \boldsymbol D\,,
\end{split}
\end{equation}
where $\boldsymbol D := \varphi^*\boldsymbol d$ is the material direction corresponding to $\boldsymbol d$. 
Written in terms of their physical components,\footnote{In a general curvilinear coordinate system, the components of a tensor field representing a given physical quantity may not necessarily carry the right physical dimension. The physical components of a tensor field do however carry the right physical dimension and can be readily interpreted as the corresponding physical quantity they represent. For a $\left({}^1_1\right)$-rank tensor $\boldsymbol T\,$, we denote its physical components as barred and are given by \citep{Truesdell1953}
\begin{equation}
\bar T^A{}_B = T^A{}_B \sqrt{G_{AA}} \sqrt{G^{BB}}\,, \quad \textrm{no summation over A nor B.}
\end{equation}
}
the radial and circumferential aforementioned quantities read
\begin{equation}\label{traction}
\begin{split}
{\boldsymbol f}_{R}(R,\Phi) =&
	\left[
	- \frac{a'(\Phi ) + (\ln (R)+1) b'(\Phi ) + \int_0^{\Phi } b(\zeta ) d\zeta}{R^2}
	- \frac{c(\Phi)}{R} \right.\\
	&\left.+\frac{h^3 \mu  (\lambda +\mu ) }{3 (\lambda +2 \mu )}\frac{\frac{1}{2}\kappa^2 (\Phi ) + \alpha }{R^2} \right] \boldsymbol u
	- \left[\frac{a(\Phi) + \ln(R)b(\Phi)}{R^2}\right] \boldsymbol t\,;\\
{\boldsymbol f}_{\Phi}(R,\Phi) =&
	- \left[\frac{a(\Phi) + \ln(R)b(\Phi)}{R^2}\right] \boldsymbol u
	+ \left[-\frac{h^3 \mu  (\lambda +\mu ) }{3 (\lambda +2 \mu )}\frac{\frac{1}{2}\kappa^2 (\Phi ) + \alpha }{R^2}\right.\\
	&\left.+\frac{\int_0^{\Phi } b(\zeta ) d\zeta}{R^2}\right] \boldsymbol t\,;\\
{\boldsymbol m}_{R}(R,\Phi) =& \frac{h^3 \mu  \lambda }{6 (\lambda +2 \mu )} \frac{\kappa (\Phi )}{R} \,\boldsymbol u \,; \quad
{\boldsymbol m}_{\Phi}(R,\Phi) = \frac{h^3 \mu  (\lambda +\mu ) }{3 (\lambda +2 \mu )} \frac{\kappa (\Phi )}{R} \,\boldsymbol t \,;\\
 \bar s_{R}(R,\Phi) =&
	\frac{h^3 \mu  (\lambda +\mu ) }{3 (\lambda +2 \mu )} \frac{\kappa (\Phi )}{R^2}\,; \quad
 \bar s_{\Phi}(R,\Phi) =
	\frac{h^3 \mu  (\lambda +\mu ) }{3 (\lambda +2 \mu )} \frac{\kappa' (\Phi )}{R^2}\,;
\end{split}
\end{equation}
where $\boldsymbol t = \varphi_*\partial_\Phi/\|\varphi_*\partial_\Phi\|$ and $\boldsymbol u = \varphi_*\partial_R\,$. Note that $\boldsymbol t$ and $\boldsymbol u$ are the unit tangent and the unit in-plane normal, respectively, of the spherical curve parametrized by Eq.~\eqref{ansatz} for a fixed $R\,$\textemdash see Fig.~\ref{not_potato}.

Our traction and out-of-plane shear results in Eq.~\eqref{traction} agree with those for d-cones that were reported by Guven and M\"uller \citep[Eqs.~19-23]{Guven2008}\textemdash note that we have additionally computed the moment vectors as well. Comparing their notation to ours: their $\boldsymbol f_{\perp}$ corresponds to our $-\boldsymbol f_{R} + \bar s_{R} \boldsymbol n\,$; their $\boldsymbol f_{\parallel}$ corresponds to our $-\boldsymbol f_{\Phi} - \bar s_{\Phi} \boldsymbol n\,$; their $C_{\parallel\perp}$ corresponds to our $a\,$; their $C_{\parallel}$ corresponds to our ${\int_0^{\Phi } b(\zeta ) d\zeta-\alpha\,}$; and their $C_\perp$ corresponds to our $c\,$\textemdash note however that they implicitly assume $h^3 \mu  (\lambda +\mu ) /\left[3 (\lambda +2 \mu )\right]$ to be equal to $1\,$.
Furthermore, if $a(\Phi)=b(\Phi)=0\,$, we obtain the same results for the traction, moment, and out-of-plane shear given in Eq.~\eqref{traction}, as those reported for d-cones by Cerda and Mahadevan \citep[\S 4.b]{cerda2005confined}. In their notation: their $a^2$ corresponds to $\alpha +1$, $B$ corresponds to $h^3 \mu (\lambda +\mu ) / \left[3 (\lambda +2 \mu )\right]\,$, and $\sigma$ corresponds to $\lambda/\left[2(\lambda+\mu)\right]\,$.

The traction-free boundary conditions on the inner and outer boundaries, i.e., $\bar{\boldsymbol f}_{R}(R_o,\Phi)=\bar{\boldsymbol f}_{R}(R_i,\Phi)=\boldsymbol 0\,$, yield that $a(\Phi)=b(\Phi)=0\,$. However, they also yield two possible expressions for $c(\Phi)$:
$$\frac{h^3 \mu  (\lambda +\mu )}{3 (\lambda +2 \mu )}\frac{\frac{1}{2}\kappa (\Phi )^2+\alpha}{R_o}\quad\textrm{or}\quad\frac{h^3 \mu  (\lambda +\mu )}{3 (\lambda +2 \mu )}\frac{\frac{1}{2}\kappa (\Phi )^2+\alpha}{R_i}\,.$$
This situation leads to a contradiction unless $\kappa$ and $\alpha$ are identically zero, which in itself is an undesirable scenario as it corresponds to a circumferentially shrinking flat disk.
Moreover, the moment ${\boldsymbol m}_{R}$ and out-of-plane shear $\bar s_{R}$ cannot satisfy the zero boundary conditions on the inner and outer boundaries either, unless $\kappa$ is identically zero.
Therefore, in order to resolve these incompatibilities, it appears necessary to introduce both an inner and an outer boundary layer as we approach $R=R_o$ and $R=R_i\,$, respectively. In these boundary layers, both the inextensibility assumption and the ansatz given in Eq.~\eqref{ansatz} would need to be relaxed for the system to be equilibrated therein. Nevertheless, the proposed model yields results that are valid far enough from the boundary and we do not concern ourselves in this work with the study of these boundary layers.

Let us now look back at the remaining governing equation, i.e., Eq.~\eqref{b_elastica}, as we have yet to solve for $\kappa\,$. 
Following the discussion above on the boundary conditions, it was established that $a(\Phi)=b(\Phi)=0\,$. Therefore, Eq.~\eqref{b_elastica} simplifies to
\begin{equation}
\label{elastica_ODE}
\kappa ''(\Phi ) + \frac{\kappa^3 (\Phi )}{2} + \left(1+\alpha \right)\kappa (\Phi ) = 0\,.
\end{equation}
Here, it is appropriate to highlight that the equation above, i.e., Eq.~\eqref{elastica_ODE}, which is describing the e-cone, may be interpreted as the spherical elastica problem~\cite{singer2008lectures}\textemdash since $\kappa$ denotes the normal curvature of a curve living on a sphere. Note that Eq.~\eqref{elastica_ODE} is identical to the what was previously reported on conical solutions in \cite{muller2008conical, efrati2015confined, cerda2005confined, guven2011conical, AnAdDi19FoldOrig}.
Further, the solution for the spherical elastica problem, as written in Eq.~\eqref{elastica_ODE}, may be explicitly given in terms of elliptic functions. In this case, it reads
\begin{equation}\label{kappa_sol}
\kappa ( \Phi ) = \kappa_o \operatorname{cn} \left[ \frac{\kappa_o}{2k} \left( \Phi - \Phi_o \right) , k \right] \,, \quad k=\frac{\kappa_o}{\sqrt{2\kappa_o^2 + 4 (1+\alpha)}}\,,
\end{equation}
where $\operatorname{cn} \left( . ,k \right)$ denotes the elliptic cosine function with elliptic modulus $k\,$. Note that the solution given by Eq.~\eqref{kappa_sol} of the spherical elastica problem, i.e., Eq.~\eqref{elastica_ODE}\textemdash which was also given in \cite{efrati2015confined, cerda2005confined}, may be alternatively written using the delta amplitude function: $\kappa ( \Phi ) = \kappa_o \operatorname{dn} \left[ \kappa_o \left( \Phi - \Phi_o \right)/2,1/k \right]$ (as in \citep{AnAdDi19FoldOrig}), or using the elliptic sine function (as in \citep{muller2008conical, guven2011conical}): $\kappa ( \Phi ) = \kappa_o \operatorname{sn} \left[\kappa_o \left( \Phi - \Phi_o \right)/(2k_1) + K(k_1),i k_1 \right] \,$, where $k_1=\kappa_o/\sqrt{\kappa_o^2 + 4 (1+\alpha)}\,$, $i$ is the imaginary unit ($i^2=-1$), and $K(.)$ is the elliptic integral of the first kind. One may prove the three different solution forms to be equal.

In order to wrap up the solution of the e-cone problem, we still have to solve for $\theta(\Phi)$ by integrating the following Ordinary Differential Equation (ODE)
\begin{equation}\label{shape_ODE}
\begin{gathered}
\frac{\cot [\theta (\Phi )] \left(1-{\theta'}^2(\Phi )\right)-\theta ''(\Phi )}{\sqrt{1-{\theta'}^2(\Phi )}}=\kappa_o \operatorname{cn} \left[ \frac{\kappa_o}{2k} \left( \Phi - \Phi_o \right) , k \right]\,,\\
{\sin\left[\theta(0)\right]\theta'(0) = \sin(\eta_1)} \,,\quad {\sin\left[\theta(2\pi)\right]\theta'(2\pi) = \sin(\eta_2)}\,,\\
\theta(0) = \theta(2\pi) = \frac{\pi}{2} \,, \quad \int_0^{2\pi} \frac{\sqrt{1-{\theta'}^2(\zeta )}}{\sin [\theta (\zeta )]} \, d\zeta = 2\pi - \psi \,,
\end{gathered}
\end{equation}
where there are five boundary conditions to solve a second order ODE in $\theta$ and find three unknown constants: $\kappa_o\,$, $k\,$, and $\Phi_0\,$\textemdash the unknown $\alpha$ follows from Eq.~\eqref{kappa_sol} and reads $\alpha = \kappa_0^2/(4k^2)-\kappa_0^2/2-1\,$. We proceed to numerically integrate Eq.~\eqref{shape_ODE} by a finite difference scheme for the ODE\textemdash using the Differential Quadrature Method on a Chebyshev grid\textemdash together with a Newton-Raphson based method to find the values of $\kappa_o\,$, $k\,$, and $\Phi_0\,$: some initial guesses are given for $\kappa_o\,$, $k\,$, and $\Phi_0\,$; the ODE is numerically integrated by finite difference for these guesses using the boundary conditions $\theta(0) = \theta(2\pi) = \pi/2$; then a Newton-Raphson based method is used to iteratively find the values of the parameters $\kappa_o\,$, $k\,$, and $\Phi_0\,$ that satisfy the remaining boundary conditions (note that the ODE is numerically integrated at each step).

\section{Discussion of the results}
\label{results}

Numerical integration of Eq.~\eqref{shape_ODE} reveals\textemdash not surprisingly\textemdash the non-uniqueness of its solution for some values of $\eta_1$ and $\eta_2\,$. Indeed, we are dealing with a second order boundary value problem coupled with three boundary conditions acting as nonlinear constraints; as such, it does not necessarily have a unique solution.
In what follows, we assume boundary conditions with mirror symmetry, i.e., $\eta=\eta_1=-\eta_2\,$, and we set out to explore the structure of the solution space of Eq.~\eqref{shape_ODE}. To do so, we track the solution orbits in an appropriately chosen phase space.
A natural choice of such a space would be $\{\eta,m_l\}$, where $m_l$ denotes the magnitude of the lips moment: the conjugate variable of the lips rotation $\eta\,$. Note however that, from Eq.~\eqref{traction}, the lips moment is proportional to the lips normal curvature $\kappa_l\,$. Thus, we equivalently choose the phase space to be $\{\eta,\kappa_l\}$. Indeed, having two lips means that $\kappa_l$ may be equal to either $\kappa(0)$ or $\kappa(2\pi)\,$. In the case of a symmetric configuration, they are identical and the orbit is represented by a single curve; otherwise, they may be different and the orbit would be represented by two simultaneous curves, one for each lip.

In the presence of different configurations for the same set of boundary conditions (same $\eta$ in the present case), barring any defects in the disk, the configuration that ought to be observed is the one with the lowest elastic energy. In order to examine the energy landscape of the solution space of Eq.~\eqref{shape_ODE}, we first compute the elastic energy of a given e-cone configuration; following Eq.~\eqref{S_V-K}, it reads
\begin{equation}\label{tot_E}
\mathcal E = \ln\left(\frac{R_o}{R_i}\right)\frac{h^3 \mu  (\lambda +\mu )}{6 (\lambda +2 \mu )}\int_{0}^{2\pi}\kappa (\zeta)^2 d\zeta\,.
\end{equation}
From this point on, we will be looking at the rescaled energy $\int_{0}^{2\pi}\kappa (\zeta)^2 d\zeta\,$ to perform such an examination. As $\eta$ varies, the family of solutions $\kappa$ of the governing ODE, i.e., Eq.~\eqref{shape_ODE}, implicitly depend on $\eta\,$. We may hence compute $\partial^2/\partial \eta^2 \left[\int_{0}^{2\pi}\kappa (\zeta)^2 d\zeta\right]$ to evaluate the stability of any given solution.

\begin{figure}[ht]
\centering
\def\svgwidth{\textwidth}
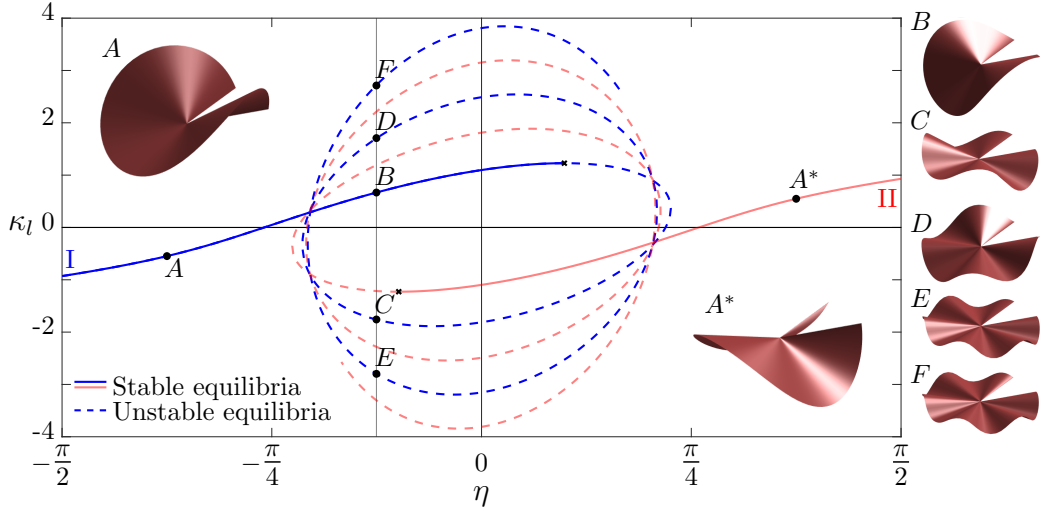
\caption{Symmetric solution orbits for $\psi=\pi/5$ in $\{\eta,\kappa_l\}$.
The orbits I and II do not connect; they are a reflection of each other with respect to the origin $\{0,0\}$; two antipodal points on either orbits correspond to configurations that mirror image each other, e.g., $A$ and $A^*\,$.
The configurations $B$-$F$ show fundamental equilibrium states of the e-cone at $\eta=-{\pi/8}$. See Mov.~1 in the supplementary material for an animated movie depicting the evolution of the e-cone configurations along the orbit I.}
\label{loop}
\end{figure}

Assuming mirror symmetric boundary conditions, we first look for symmetric solutions, i.e., solutions satisfying $\theta(\Phi)=\theta(2\pi-\Phi)$ {and $\phi(\Phi)=2\pi-\phi(2\pi-\Phi)$}. Fig.~\ref{loop} shows the truncated symmetric solution orbits in the phase space $\{\eta,\kappa_l\}$ for an excess angle $\psi = \pi/5\,$. { Separated by an $\boldsymbol{\mathsf x}$ mark on either orbits, the solid lines indicate stable equilibria while the dashed lines indicate unstable equilibria.} {Starting with vertically oriented lips, i.e., $\left|\eta\right|={\pi}/{2}\,$}, we find a unique solution on either ends, and we rotate the lips to track the symmetric equilibrium orbits {in the phase space $\{\eta,\kappa_l\}$}. We find two such orbits; and the numerical computations performed suggest they do not connect.
As a matter of fact, as shown in Fig.~\ref{loop}, there seems to be no equilibrium path of symmetric configurations leading from configuration $A$ to its mirror image, configuration $A^*\,$; instead, starting from $A\,$, the e-cone symmetric configuration adopts increasingly higher energy deformation states throughout its evolution along the orbit I. In particular, the configurations $B$-$F$ are a subset of the fundamental equilibrium states of the e-cone at $\eta=-\pi/16$ ordered alphabetically from lower to higher energy modes. See Mov.~1 in the supplementary material for an animated movie depicting this evolution.

\begin{figure}[t!]
\centering
\def\svgwidth{.85\textwidth}
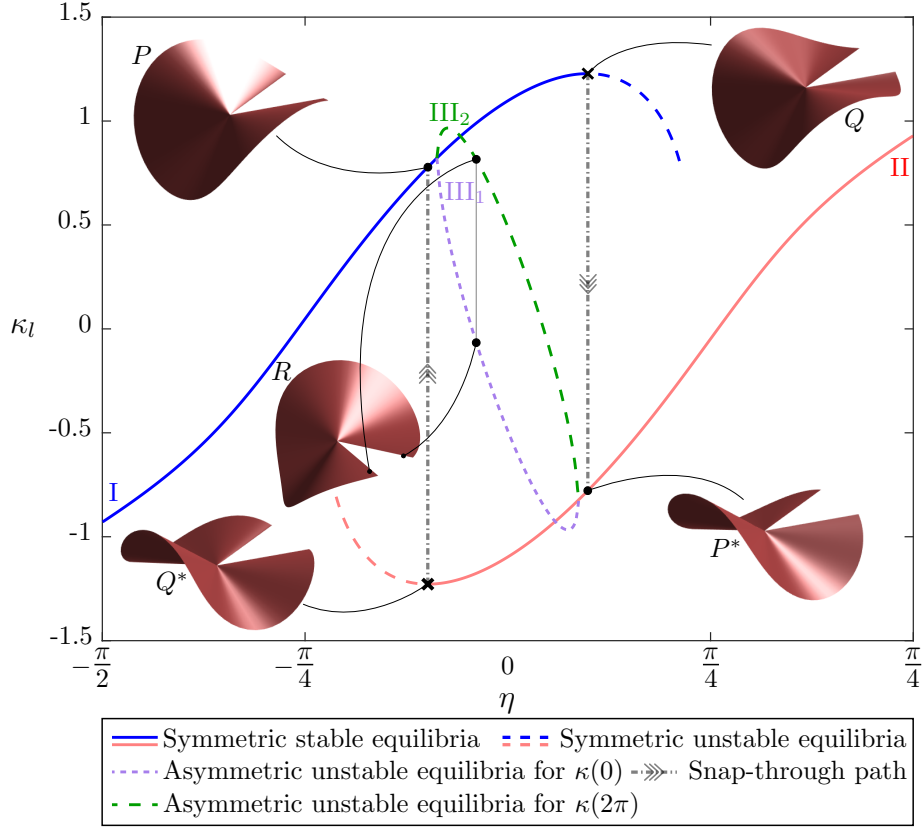
\caption{Symmetric solution orbits and the asymmetric connecting path in $\{\eta,\kappa_l\}$ for $\psi=\pi/5\,$.
The symmetric solution orbits I and II are connected via a pair of simultaneous paths $\left(\textrm{III}_1,\textrm{III}_2\right)\,$ of asymmetric unstable solutions. The asymmetric paths $\textrm{III}_1$ and $\textrm{III}_2$ each track the normal curvature of one of the lips, $\kappa(0)$ and $\kappa(2\pi)\,$, respectively, unlike the symmetric paths I and II where both lips have the same normal curvature $\kappa_l\,$. See Mov.~2~\&~3 in the supplementary material for an animated version of the figure.}
\label{orbit}
\end{figure}

\begin{figure}[t]
\centering
\def\svgwidth{.85\textwidth}
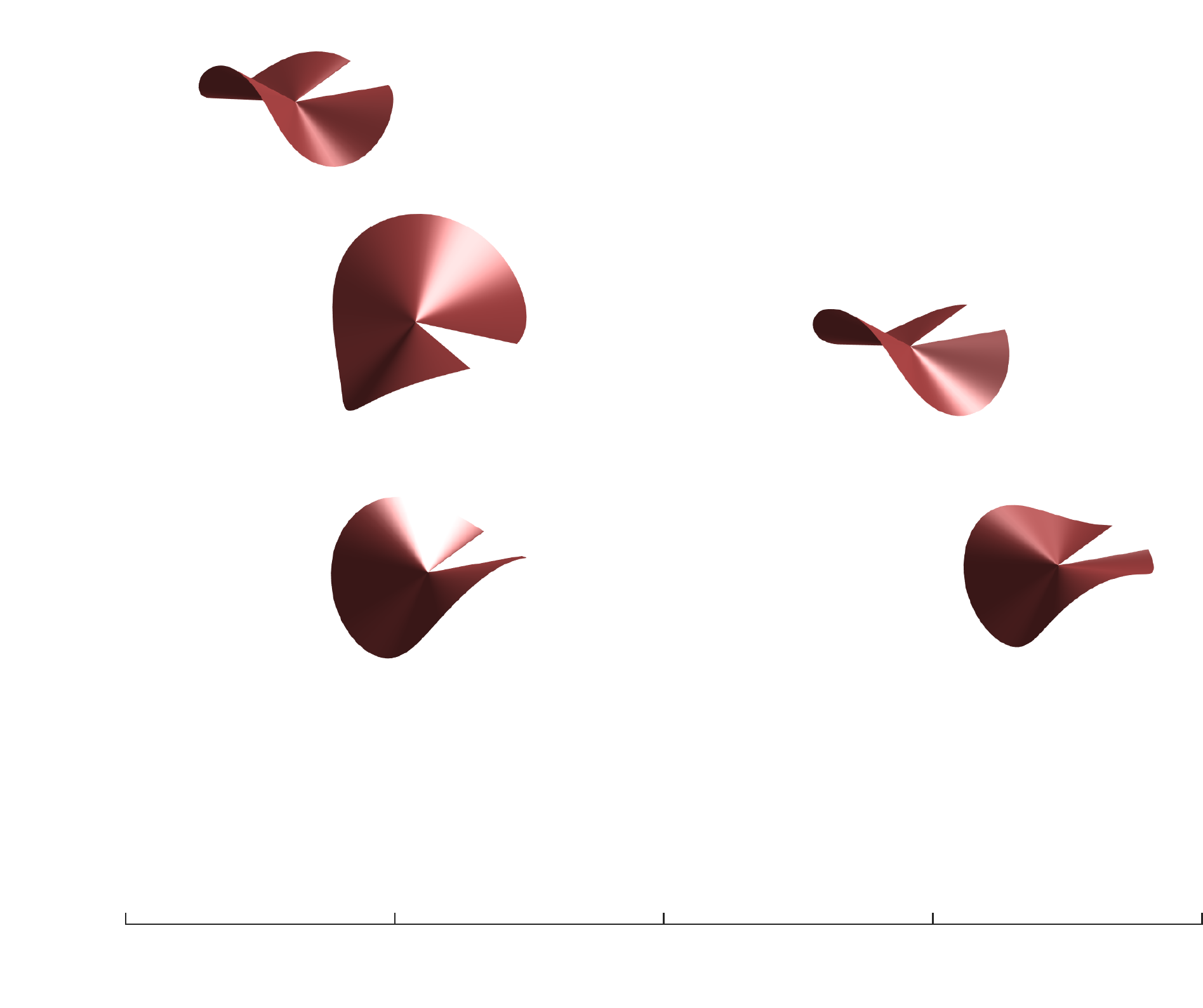
\caption{An alternative representation of the symmetric solution orbits and the asymmetric connecting path shown in Fig.~\ref{orbit}. In here, they are represented in the $\{\eta,\hat z_a\}$ phase space for $\psi=\pi/5\,$, where $\hat z_a=z_a/R_o\,$; $z_a$ being the deflection of the antipodal point to the slit as shown in the top right box.}
\label{z_a}
\end{figure}

\begin{figure}[t]
\centering
\def\svgwidth{.9\textwidth}
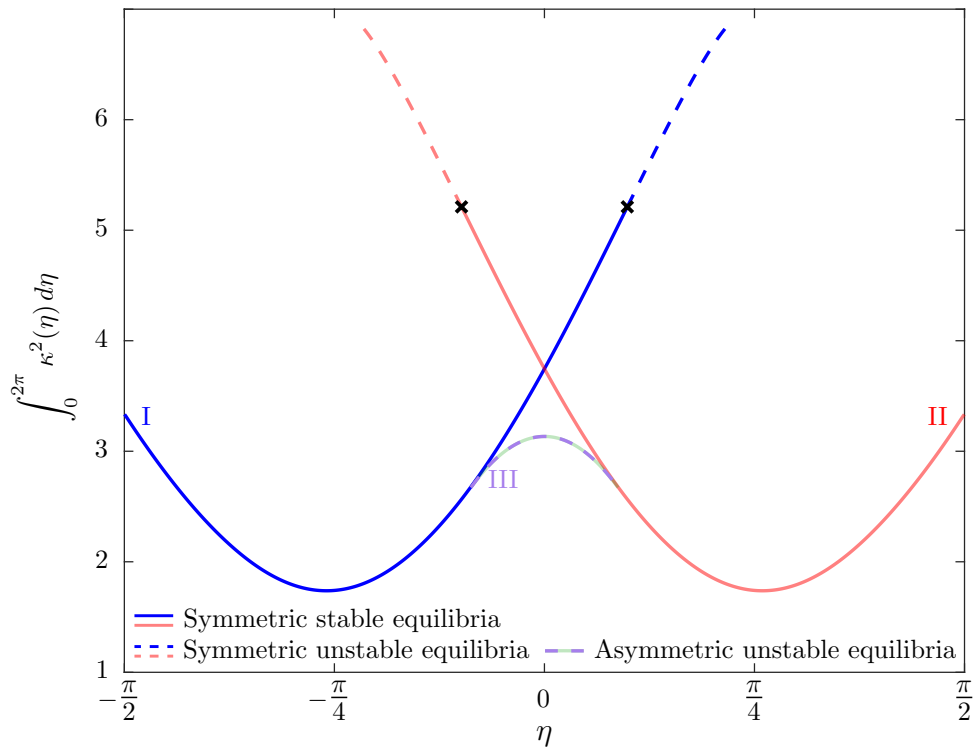
\caption{The energy landscape of the orbits I, II, and III shown in Fig.~\ref{orbit}~\&~\ref{z_a} for $\psi=\pi/5\,$. See Mov.~2 in the supplementary material for an animated version of the figure.}
\label{Energy}
\end{figure}

However, when the symmetry constraint is relaxed\textemdash allowing for the e-cone to adopt asymmetric configurations, yet still satisfying the mirror symmetric boundary conditions $\eta_1=-\eta_2\,$\textemdash we are able to track another branch consisting of asymmetric unstable equilibrium states. This reveals a path connecting the two symmetric stable solution orbit branches. Note, however, that unlike the symmetric configurations, the asymmetric configurations do not satisfy $\kappa(0) = \kappa(2\pi)\,$; the path of asymmetric unstable equilibria is hence represented in the phase space $\{\eta,\kappa_l\}$ by two simultaneous curves, one for each lip: $\textrm{III}_1$ for $\kappa(0)$ and $\textrm{III}_2$ for $\kappa(2\pi)\,$, as shown in Fig.~\ref{orbit}\textemdash see Mov.~2 in the supplementary material for an animated version. Looking for example at the configuration R, despite the mirror symmetry of the boundary conditions at its lips, i.e., $\eta_R=\eta_1=-\eta_2\,$, this configuration is asymmetric {and the normal curvatures of its lips may be respectively found where the vertical line at $\eta=\eta_R$ intersects the simultaneous curves $\textrm{III}_1$ and $\textrm{III}_2$}.
Alternatively, in Fig.~\ref{z_a}, we show the representation of the same orbits in the phase space $\{\eta,\hat z_a\}$, where $\hat z_a=z_a/R_o\,$, $z_a$ being the deflection of the antipodal point to the slit. Similarly to Fig.~\ref{orbit}, the orbit branch of asymmetric unstable solutions III connects the the two orbit branches of symmetric solutions I and II. In this case, however, we do not observe the breaking of the orbit III into two simultaneous orbits since we are tracking the deflection of a single point {(as opposed to the curvatures of the two lips)}. Moreover, we also see that the asymmetric orbit passes through the origin point $\{0,0\}$, thus highlighting the symmetry of the phase space with respect to the origin. In order to further understand the phase space orbits mapped above, we show the corresponding energy landscape in Fig.~\ref{Energy}. We plot the rescaled elastic energy of each of the configurations on the orbits I\textendash III as a function of the control parameter $\eta\,$. Note that the stability limits of the symmetric orbits correspond to the inflection points shown therein, respectively. Also, it is interesting to note that orbit III presents a lower energy alternative to both orbits I~and~II. In Mov.~2 in the supplementary material, we also show how the energy of the e-cone configurations evolves across the energy landscape as it goes from the symmetric orbit I to the symmetric orbit II via the asymmetric path III.

In both Fig.~\ref{orbit}~\&~\ref{z_a}, starting from {vertically oriented lips on the symmetric orbit I (II, respectively), i.e., $\eta=-\pi/2$, ($\eta=\pi/2$, respectively)}, the orbit describes a path of stable and symmetric configurations, e.g., configuration $P$ (configuration $P^*$, respectively) until it reaches its limit of stability, i.e., where the symmetric configuration is no longer stable{, which is indicated by a black $\boldsymbol{\mathsf x}$ mark}. Beyond the stability limit, the e-cone may snap-through at any given point from its unstable state on orbit I (II, respectively) to the corresponding stable state on orbit II (I, respectively) along {a vertical} line of equal $\eta$, e.g., snapping from configuration $Q$ to configuration $P^*$ (snapping from configuration $Q^*$ to configuration $P$, respectively)\textemdash see Mov.~3 in the supplementary material for an animated version of the snap through behaviour from orbit I to orbit II.
However, before reaching its stability limit, the e-cone encounters a branching point where a path, orbit III, of asymmetric and unstable equilibrium configurations, e.g., configuration $R$, connects the two symmetric orbits. Although this path is of unstable equilibria, it interestingly presents a lower energy alternative to the symmetric orbits as shown by the energy landscape of the solution space presented in Fig.~\ref{Energy}. This suggests that it may be possible to drive the e-cone to follow path III before reaching the limit of stability.
Nevertheless, since this branch is unstable, staying on that orbit requires fine control and additional constraints in order to prevent witnessing a snap-through transition. Indeed, unstable equilibria are notoriously hard to maintain experimentally \citep{champneys2019happy}.

\section{Concluding remarks}
\label{conclusion}

Kirigami-inspired metamaterials are on the rise and { many} of their applications are found in the realm of mechanics. It has become clear, from the short history of these materials, that an analytical grasp of their local mechanics is crucial to effectively use them in applications relating to programming both shape and effective mechanical properties. In this article, we have presented an analytical study of local kirigami mechanics { in the isometric limit}. This amounts to studying the deformation of a thin sheet with a slit, the post-buckled shape of which yields the so-called e-cone.
We have set out the problem in the framework of geometrically nonlinear plate theory and solved the outer post-buckling problem of e-cones. We found their shape as the solution of the spherical elastica problem. We also found a closed-form solution for the full stress field (traction, moment, and shear) around their apex in the case of a Saint-Venant-Kirchhoff plate model. Further, we were able to map out the full space of solutions and investigate the stability for the slit opening with mirror symmetric boundary conditions on the lips' rotations.

These solutions are valid away from the apex of the e-cone, thus providing candidates to the outer-solution of a problem manifesting a logarithmic singularity at the tip of the slit. It is noteworthy that the nature of the e-cone is such that the opening of lips can be interpreted as an edge disclination of negative charge. This in turn leads to concentration of negative Gaussian curvature, which serves as a source of stress potential~\citep{moshe2019kirigami}. We believe that this is the source of the inherent singular nature of this problem, which corresponds to the inner-solution of a boundary layer problem partially resolving the crack tip singularity. Besides, we have observed that the post-buckled solution would only be compatible with a specifically chosen set of boundary traction, moment, and shear. This suggests that it would be necessary to introduce a second boundary layer on the outer boundary of the e-cone to ensure the compatibility of the post-buckled solution with any chosen set of boundary conditions.
Although these questions go beyond the scope of this article, we foresee that our work { as} part of a more complete story. { We make further contribution towards this end in a forthcoming work based on a linearised creased model \cite{SaDias2021CreaKiri}}. We hope that, after having provided the community with the full space of post-buckling solutions corresponding to the outer-problem, more progress will be possible from past attempts in solving or offering fundamental insights to questions involving cracks on thin elastic plates and shells undergoing large out-of-plane deflections~\citep{Hui1998,Zehnder2005,Williams1961,Sih1973,Zucchini2000}. There are many applications in which having a full understanding of the post-buckling behaviour of cuts or cracks in thin structures would play a fundamental role; to name a few: aircraft fuselage fatigue under in- and out-of-plane loading~\citep{Potyondy1995,Harris1998}; and tearing of brittle sheets~\citep{Audoly2005,Bayart2010,Roman2013}.

\section*{Acknowledgements} 
The authors would like to thank the Velux Foundations for support under the Villum Experiment program (Project No. 00023059).






\end{document}